\documentclass[aip,pof,reprint,amsmath,onecolumn,bibnotes]{revtex4-1} 
\usepackage[pdftex,bookmarks,colorlinks,allcolors=blue]{hyperref}
\pdfoutput=1
\usepackage[pdftex]{graphicx}
\usepackage{natbib}
\usepackage{amsmath}
\usepackage{upgreek}
\usepackage[paperwidth=190mm,paperheight=267mm,centering,hmargin=2.5cm,vmargin=2.5cm]{geometry}
\begin{document}
	
\title{Laser impact on a drop}

\author{Alexander L. Klein} \email[]{alexludwig.klein@utwente.nl}
\author{Claas Willem Visser}
\author{Wilco Bouwhuis}
\affiliation{Physics of Fluids Group, Faculty of Science \& Technology, MESA+ Institute \& J.M. Burgerscentrum, University of Twente, P.O. Box 217, 7500 AE Enschede, The Netherlands.}
\author{Henri Lhuissier}
\affiliation{IUSTI, Aix-Marseille Universit\'e \& CNRS, 13453 Marseille Cedex 13, France.}
\author{Chao Sun}
\affiliation{Physics of Fluids Group, Faculty of Science \& Technology, MESA+ Institute \& J.M. Burgerscentrum, University of Twente, P.O. Box 217, 7500 AE Enschede, The Netherlands.}
\author{Jacco H. Snoeijer}
\affiliation{Physics of Fluids Group, Faculty of Science \& Technology, MESA+ Institute \& J.M. Burgerscentrum, University of Twente, P.O. Box 217, 7500 AE Enschede, The Netherlands.}
\affiliation{Mesoscopic Transport Phenomena, Eindhoven University of Technology, Den Dolech 2, 5612 AZ Eindhoven, The Netherlands.}
\author{Emmanuel Villermaux}
\affiliation{IRPHE, Aix-Marseille Universit\'e, 13384 Marseille Cedex 13 \&\\ Institut Universitaire de France, Paris, France.}
\author{Detlef Lohse} \email[]{d.lohse@utwente.nl}
\author{Hanneke Gelderblom} \email[]{h.gelderblom@utwente.nl}
\affiliation{Physics of Fluids Group, Faculty of Science \& Technology, MESA+ Institute \& J.M. Burgerscentrum, University of Twente, P.O. Box 217, 7500 AE Enschede, The Netherlands.}




%
%
\begin{figure}[h!]
	\includegraphics[width=1\textwidth]{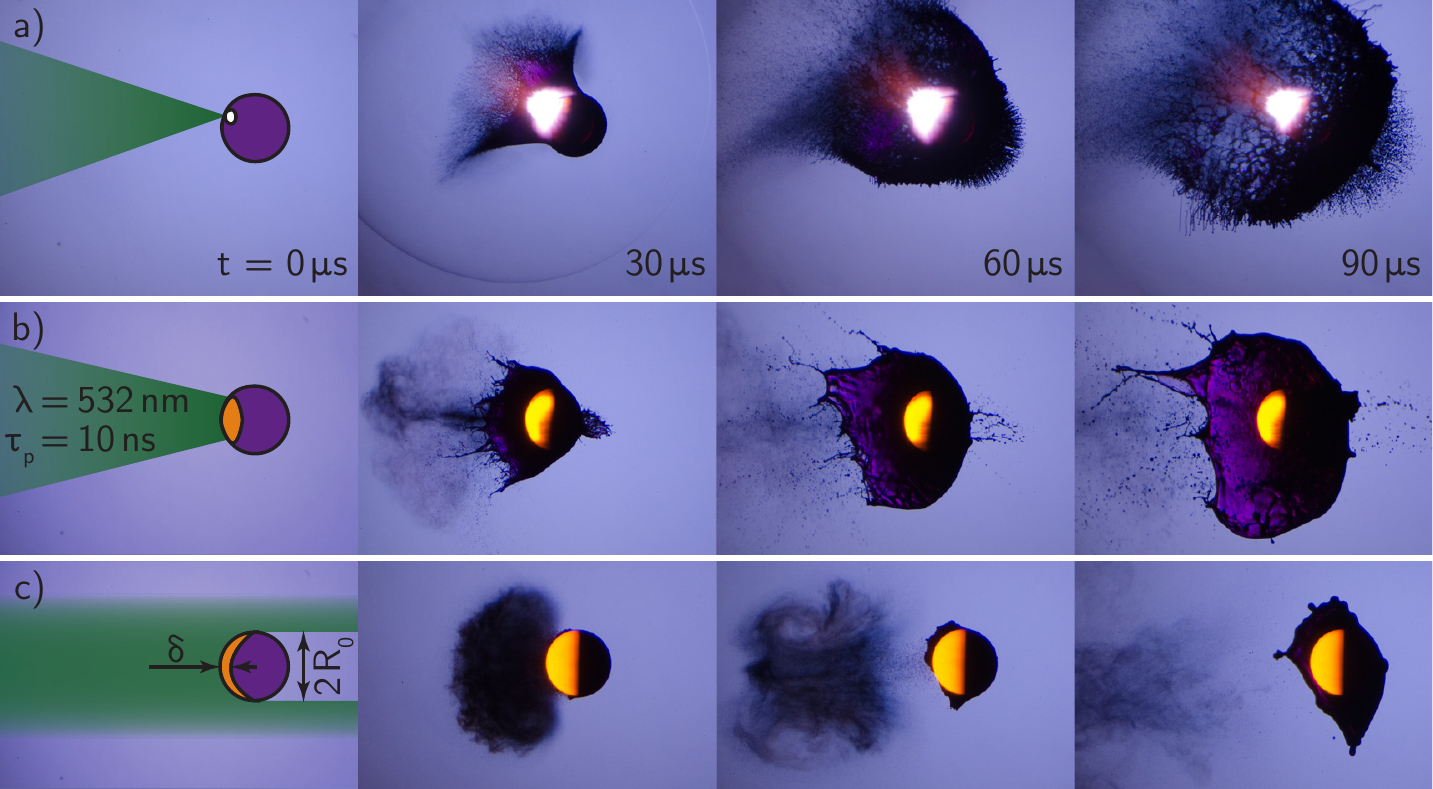}
	\caption{Laser pulses of duration $\tau_\mathrm{p}$, wavelength $\lambda$ and constant total energy impacting from the left on magenta-dyed water drops with an initial radius $R_0 = 0.9\,\mathrm{mm}$.
	The dye limits the penetration depth of the laser light $\delta$ to a superficial layer of the drop, i.e.\ $\delta/R_0 \ll 1$.
	Images are taken 30, 60 and 90 $\upmu\mathrm{s}$ after impact with a color camera and stroboscopic backlight illumination.
	The sketches in the first column illustrate the respective optical arrangements: 
	a) Tightly focused laser beam leading to a white plasma glow and a violent ablation from the drop. A spherical shockwave is visible at $t=30\,\upmu\mathrm{s}$.
	b) Moderately focused laser beam resulting in a strongly curved liquid sheet.
	c) Uniform laser irradiation ablating the drop surface and ejecting a cloud of mist in opposite direction to the laser beam. Note that the laser energy actually absorbed by the drop varies between a), b) and c).
	The shutter of the camera is open during the whole experiment to capture both, the light emitted by the plasma or fluorescent effect just after laser impact and the hydrodynamic response of the drop.
	 Source: APS-DFD (\url{http://dx.doi.org/10.1103/APS.DFD.2014.GFM.V0016})}
	\label{fig:LaserImpactColor} 
\end{figure}
\maketitle

The impact of a laser pulse on a highly-absorbing liquid drop can lead to a violent response: the drop is accelerated, strongly deforms, and eventually fragments.
Shock waves, the ejection of matter, and even plasma formation can accompany this process (see Fig.~\ref{fig:LaserImpactColor})

The total energy absorbed by the drop and its spatial distribution in the superficial layer determine the hydrodynamic response to the laser impact.
For a localized energy deposition the threshold of optical breakdown in water is exceeded, which provokes a violent drop explosion (Fig.~\ref{fig:LaserImpactColor}a).
When the energy density in the superficial layer is decreased, plasma generation is avoided and a fluorescence effect shifting the green laser light to yellow becomes visible (Fig.~\ref{fig:LaserImpactColor}a, b).
The deformation of the drop changes with the spatial distribution of the absorbed energy: from a strongly curved thin liquid sheet in Fig.~\ref{fig:LaserImpactColor}b to a flatter liquid sheet in Fig.~\ref{fig:LaserImpactColor}c.

\begin{figure*}
	\includegraphics{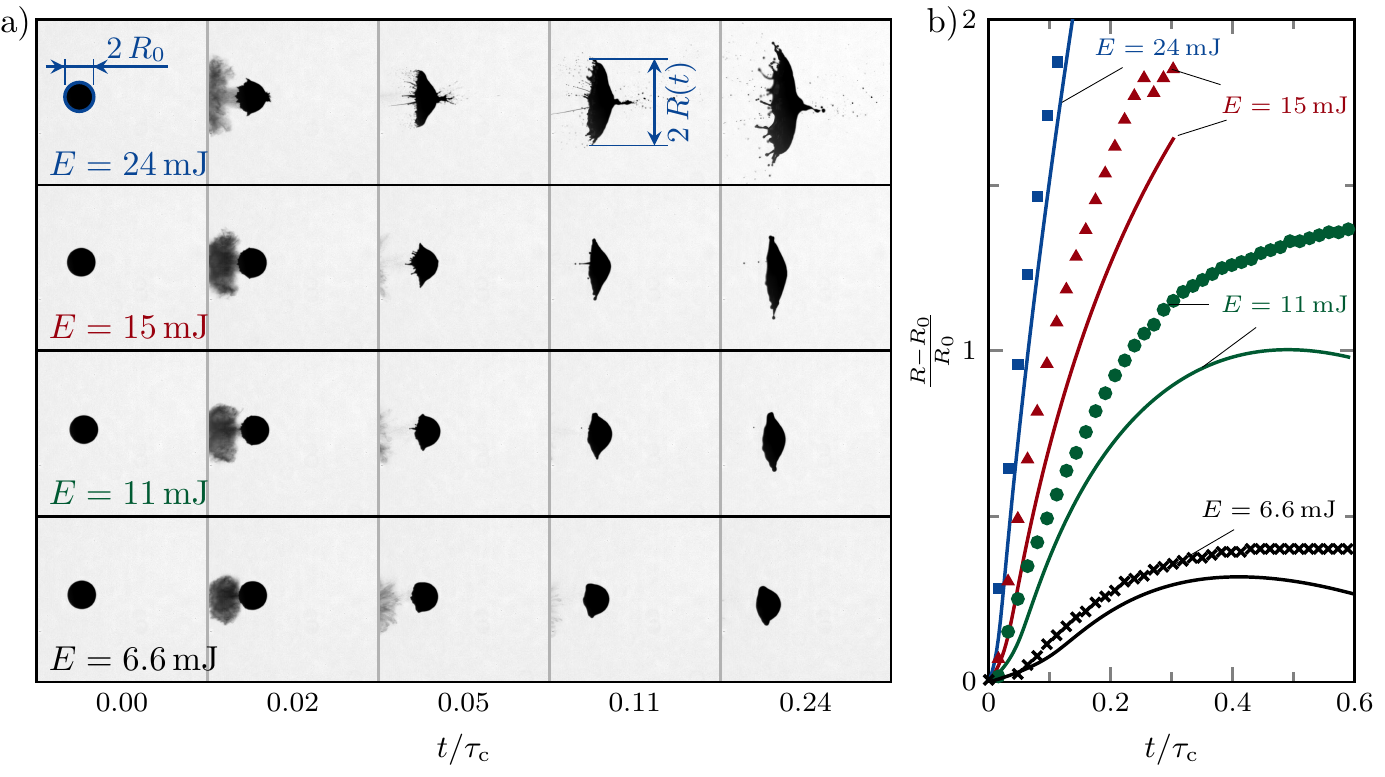}
	\caption{(Color online) Drop shape dynamics for an uniform laser irradiation (configuration Fig.~\ref{fig:LaserImpactColor}c) with different laser-pulse energies.
	a) High-speed images for a dyed water drop with initial radius $R_0 = 0.9\,\mathrm{mm}$ hit by a laser pulse at $t = 0$.
	$E$ is the energy that is absorbed by the drop (increasing from bottom to top).
	b) Comparison between experiments (markers) and BI simulations (solid lines) for the radial expansion of the drop for four different laser energies.}
	\label{fig:HighSpeed} 
\end{figure*}

The deformation of the drop occurs on the inertial time-scale $\tau_{\rm i} = R_0/U \sim 10^{-4}$ to $10^{-3}\,\mathrm{s}$, with $U$ being the propulsion speed of the drop, and is eventually slowed down by surface tension $\gamma$ on the capillary time-scale $\tau_\mathrm{c}=\sqrt{\rho R_0^3/\gamma} =3.5\,\mathrm{ms}$ (see Fig.~\ref{fig:HighSpeed}). 
Both time-scales are clearly separated from those of the laser-matter interaction, namely, the laser pulse duration $\tau_p = 10\,\mathrm{ns}$ and the duration of the ejection of matter $\tau_{\rm e} \sim 10^{-5}\,\mathrm{s}$.

The separation of time-scales requires an elaborate combination of stroboscopic (Fig.~\ref{fig:LaserImpactColor}) and high-speed (Fig.~\ref{fig:HighSpeed}a) imaging techniques to resolve the physical phenomena experimentally.
However, it eases the modeling of the fluid dynamics: the complex laser-matter interaction can be replaced by an appropriate pressure pulse acting on the spherical drop \cite{Klein:2015a}.
As an example, Fig.~\ref{fig:HighSpeed}b shows the comparison between experiments and boundary-integral (BI) simulations for the deformation of the drop perpendicular to the laser pulse.

This work is part of an Industrial Partnership Programme of the Foundation for Fundamental Research on Matter (FOM), which is financially supported by the Netherlands Organization for Scientific Research (NWO).
This research programme is co-financed by ASML. W.B.~and J.H.S.~acknowledge support from NWO through VIDI Grant No.~11304.
%
%

\bibliography{main}

\end{document}